# Magnified Imaging Based on Non-Hermitian Nonlocal Cylindrical Metasurfaces

Silvio Savoia[1,2], Constantinos A. Valagiannopoulos[3], Francesco Monticone[2,4], Giuseppe Castaldi[1], Vincenzo Galdi[1], and Andrea Alù[2]

[1] *Waves Group, Department of Engineering, University of Sannio, I-82100 Benevento, Italy*

[2] *Department of Electrical and Computer Engineering, The University of Texas at Austin, Austin, Texas 78712, USA*

[3] *Department of Physics, School of Science and Technology, Nazarbayev University, KZ-010000, Astana, Kazakhstan*

[4] *School of Electrical and Computer Engineering, Cornell University, Ithaca, New York 14853, USA*

Abstract

We show that a cylindrical lensing system composed of two metasurfaces with suitably tailored non-Hermitian (i.e., with distributed gain and loss) and nonlocal (i.e., spatially dispersive) properties can perform magnified imaging with reduced aberrations. More specifically, we analytically derive the idealized surface-impedance values that are required for "perfect" magnification and imaging, and elucidate the role and implications of non-Hermiticity and nonlocality in terms of spatial resolution and practical implementation. For a basic demonstration, we explore some proof-of-principle quasi-local and multilayered implementations, and independently validate the outcomes via full-wave numerical simulations. We also show that the metasurface frequency-dispersion laws can be chosen so as to ensure unconditional stability with respect to arbitrary temporal excitations. These results, which extend previous studies on planar configurations, may open intriguing venues in the design of metastructures for field imaging and processing.

## I. INTRODUCTION

Optical metamaterials, artificially engineered so as to exhibit desired responses not readily attainable in nature, have been the subject of intense investigations over the past decades [1,2], with promises to overcome some of the fundamental limitations of optical instruments [3]. For instance, in a seminal work by Pendry [4], it was shown that a slab with negative refractive index can create a 2D image with a spatial resolution that is not bounded by the conventional diffraction limit [3]. The basic idea can be generalized to cylindrical lenses, in order to achieve image magnification [5,6]. The enabling concept for such "perfect lenses" is the possibility to recover the subwavelength information encoded in the evanescent waves, which is typically lost at wavelength-sized distances from the source. In the quasi-electrostatic regime, for a given polarization, a plasmonic material exhibiting negative permittivity (e.g., a noble metal at optical wavelengths) is sufficient to attain "superlensing" effects [4]. In the dynamic case, the negative-refractive-index requirement can be met by metamaterials engineered in various ways [7-9] depending on the operational frequency of interest.

In the superlensing effect, the evanescent-wave enhancement relies on the excitation of surface plasmons, and hence the possible applications are limited to near-field effects. To achieve *far-field* subwavelength imaging, the "hyperlens" concept was put forward [10,11], which relies on *hyperbolic* metamaterials [12] capable of transforming the (otherwise evanescent) high-transverse-wavenumber components into propagating waves inside the lens. In conjunction with a suitably shaped (e.g., obliquely cut or curved) output surface, this makes it possible to attain far-field imaging with subdiffractive resolution and magnification.

In spite of the promising theoretical predictions, experimental demonstrations of superlenses [13,14] and hyperlenses [15-17], have evidenced the inherent practical limitations, and most notably the detrimental influence of material losses [18,19]. On the other hand, alternative low-loss implementations of superlenses, e.g., based on photonic crystals [20-23], exhibit inherent resolution limitations due to the crystal-lattice periodicity. Against this background, the idea of exploiting material constituents featuring optical *gain* has recently emerged as a viable route to engineer effectively lossless metamaterials [24,25]. For instance, in [26], a gain-assisted hybrid superlens-metalens was proposed, and subdiffraction spatial resolution was numerically demonstrated.

Within this framework, the emerging field of *non-Hermitian* optics, inspired by the *parity-time* (PT) symmetry concept in quantum mechanics [27-29], has shaken the conventional wisdom of gain-induced loss compensation, opening entirely new, and largely unexplored, perspectives in the physics and engineering of gain-loss interactions. The reader is referred to [30] for a recent review of PT-symmetry in optics, and to [31-49] for a sparse sampling of studies on the implications and applications of non-Hermitian optics, including metamaterials, plasmonics and lasers, just to mention a few. In particular, of special interest for the present study is a series of recent investigations [45,48-51] on the negative-refraction, focusing, and imaging effects achievable by means of PT-symmetric metasurface pairs. In particular, in [49], it was demonstrated analytically and numerically that a pair of planar metasurfaces featuring balanced loss and gain and tailored *nonlocal* properties can act as a transversely invariant, planar lens with the potential to perform volume-to-volume imaging, with reduced aberrations and loss-sensitivity. In such a system, gain and loss do not merely compensate, but rather interplay in an anomalous fashion. More specifically, the passive metasurface is tailored to act as an omnidirectional coherent perfect absorber, while the active one acts as an omnidirectional coherent emitter. While this general concept looks attractive and promising, the inherent *afocal* character of the assumed planar configuration prevents the possibility to perform image magnification, which may be desirable in many application scenarios.

To overcome the above limitation, the present study explores the imaging capabilities of non-Hermitian *cylindrical* concentric metasurfaces. More specifically, in Sec. II, we describe the problem geometry and outline its formulation. In Sec. III, we analytically derive the ideal properties of the metasurfaces that are required to attain "perfect" cylindrical imaging. Moreover, we illustrate the non-Hermitian and nonlocal properties of the metasurfaces, and address some issues concerning the attainable spatial resolution. In Sec. IV, we explore possible implementations, numerically validate our theoretical predictions, and address stability- and implementation-related issues. Finally, in Sec. V, we discuss implications and perspectives of our results.

## II. GEOMETRY OF THE PROBLEM

Referring to the schematic in Figure 1, we consider a cylindrical geometry embedded in vacuum, infinitely long and invariant along the *z*-direction of the associated coordinate system $(r,\phi,z)$. We consider a "source" (virtual) surface at $r = R_s$ where a transversely magnetic polarized, time-harmonic field distribution [with implied $\exp(-i\omega t)$ time-dependence] is assigned in terms of the *z*-directed electric field

$$E_z(R_s, \phi) \equiv E_{sz}(\phi). \tag{1}$$

As anticipated (see also the schematic in Figure 1), we are interested in reproducing this field distribution at an "image" (virtual) surface located at $r = R_i > R_s$, so as to attain an inherent geometrical magnification by a factor $R_i/R_s > 1$, without monochromatic aberrations. By letting

$$E_z(R_i,\phi) \equiv E_{iz}(\phi) \tag{2}$$

the field distribution at the image surface, our problem can be stated in mathematical terms as attaining the condition

$$E_{iz}(\phi) = \alpha E_{sz}(\phi), \tag{3}$$

with $\alpha$ denoting a real-valued constant, henceforth assumed as $\alpha = 1$.

To realize the magnified-imaging condition implied by (3), we consider a cylindrical lensing system composed of two idealized (zero-thickness) concentric metasurfaces placed at $r = R_1$ and $r = R_2 > R_1$, with homogeneous (i.e., $\phi$-independent) surface impedances $Z_1$ and $Z_2$, respectively. While this system may appear, at first glance, as a direct generalization of the planar lens considered in [49], there are some important caveats to consider. Most notably, although the PT-symmetry condition assumed in [49] (with the two planar metasurfaces characterized by balanced positive and negative resistances) admits some possible generalizations [52,53] to cylindrical scenarios, these are not apt for our metasurface-based formulation and for the magnification goal at hand. Accordingly, although we intuitively expect a non-Hermitian and nonlocal character for the required metasurfaces, we do not make any prior assumption on their nature. Instead, we analytically derive the general conditions that they need to satisfy in order to realize ideal magnification as described by the condition in (3).

### III. THEORY AND DESIGN IMPLICATIONS

#### A. Analytical derivations

In each of the vacuum regions of interest (Figure 1), the *z*-directed electric field can be represented in terms of a Fourier-Bessel series expansion [54]

$$E_z(r,\phi) = \sum_{n=-\infty}^{\infty} \left[ A_\nu^{(n)} H_n^{(1)}(k_0 r) + B_\nu^{(n)} H_n^{(2)}(k_0 r) \right] \exp(in\phi), \quad R_{\nu-1} < r < R_\nu, \qquad (4)$$

where $\nu = 1,2,3$, and we have defined $R_0 \equiv R_s$ and $R_3 \equiv \infty$ for notational compactness. Moreover, $A_\nu^{(n)}$ and $B_\nu^{(n)}$ are sets of unknown expansion coefficients, $H_n^{(1)}$ and $H_n^{(2)}$ denote the $n$th-order Hankel function of first and second kind [55], respectively, and $k_0 = \omega/c = 2\pi/\lambda_0$ is the vacuum wavenumber (with $c$ and $\lambda_0$ denoting the corresponding speed of light and wavelength, respectively). From (4), the corresponding tangential magnetic field follows from the relevant Maxwell's curl equation

$$H_\phi(r,\phi) = \frac{i}{k_0 \eta_0} \frac{\partial E_z(r,\phi)}{\partial r}, \qquad (5)$$

with $\eta_0 = \sqrt{\mu_0/\varepsilon_0} \approx 377\Omega$ denoting the vacuum characteristic impedance. By enforcing impedance matching (i.e., zero reflection in the region $R_s < r < R_1$), so that the source signal is not perturbed, and the radiation condition (for $r > R_2$), it readily follows that

$$B_1^{(n)} = B_3^{(n)} = 0. \qquad (6)$$

Moreover, by particularizing the series expansions in (4) at the source ($r = R_s$) and image ($r = R_i$) surfaces, we obtain

$$E_{sz}(\phi) = E_z(R_s,\phi) = \sum_{n=-\infty}^{\infty} A_1^{(n)} H_n^{(1)}(k_0 R_s) \exp(in\phi), \qquad (7)$$

$$E_{iz}(\phi) = E_z(R_i,\phi) = \sum_{n=-\infty}^{\infty} A_3^{(n)} H_n^{(1)}(k_0 R_i) \exp(in\phi), \qquad (8)$$

respectively, which directly relate the sets of expansion coefficients $A_1^{(n)}$ and $A_3^{(n)}$ to the Fourier coefficients of the source- and image-field distributions, respectively. By enforcing the magnified-imaging condition (3) (with $\alpha = 1$), we therefore obtain

$$A_3^{(n)} = A_1^{(n)} \frac{H_n^{(1)}(k_0 R_s)}{H_n^{(1)}(k_0 R_i)}. \tag{9}$$

The remaining sets of unknown expansion coefficients need to be calculated by enforcing the electric-field continuity and impedance boundary conditions at the metasurfaces,

$$E_z(R_v^-, \phi) = E_z(R_v^+, \phi), \quad v = 1, 2, \tag{10}$$

$$H_\phi(R_v^+, \phi) - H_\phi(R_v^-, \phi) = \frac{E_z(R_v, \phi)}{Z_v}, \quad v = 1, 2, \tag{11}$$

where the superscripts "$-$" and "$+$" denote the conventional one-sided limits. This yields four (countably infinite) sets of linear equations, with the unknowns $A_2^{(n)}$, $B_2^{(n)}$ and the surface impedances $Z_1$ and $Z_2$. It becomes apparent that for *local* metasurfaces (i.e., $Z_1$ and $Z_2$ independent of the angular-momentum order $n$), the overall system of equations is inherently *overdetermined*, and can only be solved in a weak (e.g., least-square) sense. On the other hand, by assuming $n$-dependent surface impedances (i.e., *nonlocal* metasurfaces), the system can be solved analytically in closed form, and we obtain

$$Z_1^{(n)} = -S_n \frac{k_0 \eta_0 \pi R_1}{4} \frac{H_n^{(1)}(k_0 R_1) H_n^{(1)}(k_0 R_i)}{H_n^{(1)}(k_0 R_2)}, \tag{12}$$

$$Z_2^{(n)} = S_n \frac{k_0 \eta_0 \pi R_2}{4} \frac{H_n^{(1)}(k_0 R_2) H_n^{(1)}(k_0 R_s)}{H_n^{(1)}(k_0 R_1)}, \tag{13}$$

where the superscript "$^{(n)}$" has been added to highlight the nonlocal character, i.e., the fact that the metasurfaces present an impedance that changes with the momentum of the impinging harmonic, and

$$S_n = \frac{H_n^{(1)}(k_0 R_1) H_n^{(2)}(k_0 R_2) - H_n^{(2)}(k_0 R_1) H_n^{(1)}(k_0 R_2)}{H_n^{(1)}(k_0 R_i) - H_n^{(1)}(k_0 R_s)}. \quad (14)$$

By recalling the symmetry properties of the Hankel functions with respect to their order [55], it can be verified that

$$Z_v^{(-n)} = Z_v^{(n)}, \quad v = 1, 2. \quad (15)$$

**B. Nonlocality, non-Hermiticity and resolution issues**

Although it is evident from (13) and (14) that our cylindrical scenario inherently requires *complex-valued* impedances, it is not straightforward to analytically ascertain the non-Hermitian requirements in terms of gain and loss distribution, and to assess the actual degree of required nonlocality. Accordingly, in what follows we illustrate these effects and their implications on the imaging capabilities, by exploring representative numerical examples.

We start by considering a configuration with source and image surfaces at $k_0 R_s = 10$ and $k_0 R_i = 20$, respectively (i.e., a geometrical magnification factor $R_i / R_s = 2$), and metasurfaces at $k_0 R_1 = 13$ and $k_0 R_2 = 17$. Figure 2 shows the required surface-impedance values [from (13) and (14)] pertaining to the first 31 angular-momentum orders [in view of the symmetry condition (15), only $n \geq 0$-orders are displayed]. We observe that the real parts can assume both negative and positive values (i.e., gain and loss), thereby confirming the expected non-Hermitian character. However, unlike the planar case [49], there is no clear balance and symmetry between the inner and outer surface. In fact, due to the nonlocality, it is generally not even possible to associate to the two impedances a clear-cut active or passive character (see, e.g., the sign inversions in the impedance real-parts occurring around the modal order $n=10$). Moreover, the nonlocal character appears very pronounced in some regions, and somewhat milder in others. To gain some physical insights into this behavior, it is worth recalling that (see Sec. 9.3.1 in [11,55]) the generic *n*-th order angular-momentum mode in the Fourier-Bessel field expansion (4) decays as $\sim (k_0 r/n)^{-n}$ for

$$n \gtrsim k_0 r, \tag{16}$$

thereby implying that the cylindrical surface $k_0 r = n$ effectively represents a *caustic*, inside which the mode is essentially *evanescent*. The different color shadings in Figure 2 identify three regions across the two relevant caustics at $n = k_0 R_s$ (source) and $n = k_0 R_i$ (image). More specifically, the purple-shaded regions ($n < k_0 R_s$) contain the angular-momentum modes that exhibit a propagating character within the entire lens domain ($k_0 R_s < r < k_0 R_i$). For these modal orders, the nonlocal character does not appear very pronounced, and the two surface impedances exhibit a clear-cut passive ($Z_1$) or active ($Z_2$) behavior, with real-parts that are (in absolute value) fractions of the vacuum characteristic impedances. Conversely, in the cyan-shaded regions ($k_0 R_s < n < k_0 R_i$), containing modal orders that exhibit a caustic between the source and image surfaces, nonlocality is significantly more pronounced, with wider dynamics and faster variations. In this case, the real part of the surface impedances can change sign, thereby implying gain at some modal orders and loss at others. Finally, the orange-shaded region ($n > k_0 R_i$) contains modal orders that are evanescent within the entire lens domain. In this case, we observe vanishingly small real parts of the surface impedances, and asymptotically decreasing imaginary parts. As a matter of fact, by exploiting in (12) and (13) the large-order asymptotic expansion of the Hankel functions [55], and retaining the dominant terms, it can be shown that the surface impedances pertaining to these higher-order modes behave as

$$Z_1^{(n)} \sim -\frac{i\eta_0}{2n} k_0 R_1 \left( \frac{R_s R_2^2}{R_i R_1^2} \right)^n, \tag{17}$$

$$Z_2^{(n)} \sim i \frac{\eta_0 k_0 R_2}{2n}, \tag{18}$$

thereby confirming the essentially reactive character observed in Figure 2.

For a deeper understanding, Figure 3 shows the field (magnitude) radial distributions pertaining to three representative angular-momentum modal orders, for the parameter configuration as in Figure 2. For a low-order mode [$n=2$, Figure 3(a)], which is propagating everywhere, the two non-Hermitian surface impedances act as an open resonating cavity (with a visible standing-wave pattern), which essentially compensates the cylindrical spreading, so as to recover at the image surface the original amplitude (and phase, not shown) enforced at the source surface. For a moderately-high order [$n=15$, Figure 3(b)] mode, whose caustic is located within the lens domain, the decaying field is significantly amplified in the cavity region between the two non-Hermitian surface impedances, and then propagates to the image surface. For a higher-order [$n=30$, Figure 3(c)] mode, which is everywhere evanescent, the amplification effects in the cavity region are even more dramatic (note the semi-log scale in the graph), even though in this case the two metasurfaces are essentially reactive.

From the above discussion, it emerges that a *perfect* reconstruction at the image surface of the source-field distribution, including possible *subwavelength* details transported by high-order angular-momentum modes, would require a precise tailoring of the non-Hermitian and nonlocal response of the two metasurfaces that appears to be beyond the current and near-future technological capabilities. Within this framework, it should also be highlighted that our assumption to enforce an *arbitrary* source-field distribution is highly idealized. In practice, if we realistically assume that the source field distribution at $r = R_s$ is generated by finite-energy current sources contained in the inner cylindrical region $r < R_s$ filled by a conventional dielectric medium, the number of *degrees of freedom* [i.e., significantly nonzero $A_1^{(n)}$ coefficients in the Fourier series (7)] is inherently limited by the low-pass character of the propagation operator (see, e.g., the discussion in [56]). Remarkably, if we neglect the moderate-to-high-order angular-momentum modes (cyan- and orange-shaded regions in Figure 2), and focus on the modal orders that exhibit a propagating character within the entire lens domain (purple-shaded regions in Figure 2), the arising metasurface synthesis turns out to be significantly more approachable, since the corresponding surface impedances exhibit a milder nonlocality and an unambiguous (active or passive) character. Such operational scenario resembles the one considered for the planar case [49], in terms of *diffraction-limited* imaging and implementation complexity, but it adds the geometrical-magnification capability. Also in that scenario, in fact, the evanescent contribution to the image was neglected, and its reconstruction would have required super-oscillatory reactive surfaces for the large transverse wavenumbers associated with the evanescent spectrum of the spatial distribution to be imaged.

In what follows, with reference to diffraction-limited magnified imaging, we explore possible implementation strategies for the nonlocal cylindrical metasurfaces.

## IV. REPRESENTATIVE RESULTS

### A. Quasi-local implementation

The possibly simplest strategy to cope with the inherent nonlocal character of the impedance surfaces is to somehow mitigate the degree of nonlocality, so that *local* metasurfaces can be utilized. To illustrate this concept, we define a "nonlocality indicator"

$$F(k_0 R_1, k_0 R_2) = \frac{1}{2N(N+1)} \sum_{\substack{n,m=0 \\ m>n}}^{N} \left\{ \frac{|Z_1^{(n)} - Z_1^{(m)}|^2}{|Z_1^{(n)}|^2 + |Z_1^{(m)}|^2} + \frac{|Z_2^{(n)} - Z_2^{(m)}|^2}{|Z_2^{(n)}|^2 + |Z_2^{(m)}|^2} \right\}, \quad (19)$$

which quantifies the degree of nonlocality as a nondimensional parameter ranging from zero (locality) to one (extreme nonlocality), as a function of the metasurface positions, for given electrical radii of the source and image surfaces. Figure 4 shows the nonlocality indicator (on a dB scale) for the previously considered parameters $k_0 R_s = 10$ and $k_0 R_i = 20$, a maximum angular-momentum modal order $N = 5$, and with $k_0 R_1$ and $k_0 R_2$ spanning geometrically feasible ranges. We observe a three-order-of-magnitude dynamic range, with alternating minima and maxima for this nonlocality measure. In particular, we identify a specific parameter configuration ($k_0 \hat{R}_1 = 11.64, k_0 \hat{R}_2 = 16.43$, marked with a cyan cross in the figure), for which the indicator is as small as $-35$dB, thereby indicating a particularly mild nonlocality. For this configuration, a *local* approximation of the surface impedances in terms of average values,

$$\bar{Z}_\nu = \frac{1}{N+1} \sum_{n=0}^{N} \hat{Z}_\nu^{(n)}, \quad \nu = 1, 2, \quad (20)$$

may provide acceptably good results. Here and henceforth, the caret and overline are utilized to indicate the "optimal" parameters and their average values, respectively.

Figure 5(a) and Figure 5(b) show the ideal surface impedances for the modal orders up to $N = 5$ for the case at hand. As expected, the variations are rather mild especially in the real parts of the impedances, which are much larger than the corresponding imaginary parts. For this case, the local approximation in (20) yields $\bar{Z}_1 = (0.418 + i0.102)\eta_0$ and $\bar{Z}_2 = (-0.599 + i0.109)\eta_0$. By comparison with the planar scenario in [49], we observe that also in our case the surface impedances exhibit gain and loss. However, there is no apparent symmetry between gain and loss, and a reactive (capacitive) part is also present; these differences can be expected, and attributed to the cylindrical spreading of the wave, which breaks the PT symmetry of the desired field distribution in the planar scenario (the fact that one metasurface is entirely contained into the other rules out the position requirement for PT-symmetry). From the implementation viewpoint, similar considerations as in [49] hold. At microwave frequencies, the required gain may be attained by exploiting classical amplification schemes based on operational amplifiers and Gunn diodes [57-59]; at optical frequencies semiconductor-based active media [60-63] or parametric effects may be exploited.

To test the magnified-imaging capabilities, we consider a diffraction-limited, real-valued source-field distribution [see (7)] with constant coefficients ($A_1^{(n)} = 1$ for $n = -5,..,5$, and $A_1^{(n)} = 0$ otherwise). The corresponding field profile is shown (red-dashed curves) in Figure 5(c) and Figure 5(d) (real and imaginary parts, respectively), and is compared with the imaged field profile (blue-solid curves) obtained by exploiting the local approximation above, and calculated via the Fourier-Bessel series expansion in (4). To facilitate the comparison, the field profiles are plotted as a function of the angle $\phi$; however, a geometrical-magnification factor $R_i/R_s = 2$ needs to be accounted for at the image surface. The (dominant) real parts of the source and imaged profiles are in excellent agreement, while there is a slight residual imaginary part in the imaged profile (about an order of magnitude smaller than the real-part peak value) attributable to the local approximation. Also shown, as a reference (magenta-dotted curves) is the field profile at the image surface in the absence of the cylindrical lens ($\bar{Z}_{1,2} \to \infty$). It is clear that, in this case, results would be drastically different.

For the same parameter configuration, Figure 6 shows the (real-part) field distributions over the entire cylindrical lens domain, in the presence [Figure 6(a)] and absence [Figure 6(b)] of the metasurfaces. As already illustrated in Figure 3(a) with reference to a generic propagating mode, the two metasurfaces act as an open cavity system which corrects the propagation-induced distortions, and re-creates at the image surface a geometrically-magnified version of the enforced source-field profile.

As a further representative example Figure 7 shows the results pertaining to another parameter configuration ($k_0 R_s = 13, k_0 R_i = 18, k_0 R_1 = 13.40, k_0 R_2 = 15.65$), identified via a parametric study. By comparison with the previous example, this configuration features a smaller magnification factor $R_i/R_s = 1.38$, and a weaker nonlocality, as clearly observable from the surface impedances [Figure 7(a) and Figure 7(b)] and witnessed by the value of the nonlocality indicator ($F \approx -50\text{dB}$). In this case, the local approximation in (20) yields $\bar{Z}_1 = (0.557 + i0.205)\eta_0$ and $\bar{Z}_2 = (-0.690 + i0.103)\eta_0$, and we select a diffraction-limited complex-valued source-field profile (with coefficients given in the caption). As a consequence of the particularly weak nonlocality, the source and imaged field profiles are now in excellent agreement for both the real [Figure 7(c)] and imaginary [Figure 7(d)] parts. Once again, this is in stark contrast with the results that would be obtained the absence of the lens.

Overall, the above results indicate that, within suitable parameter ranges, non-Hermitian, local cylindrical metasurfaces can provide magnified imaging with reasonably small aberrations. Remarkably, this leads to particularly simple implementations of the required metasurfaces, in terms of thin cylindrical layers of homogeneous, isotropic materials featuring loss or gain.

**B. Multilayered implementation**

It is evident from Figure 4 that nonlocality is generally *non-negligible*, and therefore the quasi-local approach is not necessarily applicable for arbitrary scenarios. The synthesis of metamaterials and metasurfaces with tailored nonlocal (i.e., spatially dispersive) responses has recently received considerable attention in view of its increasing relevance in several application scenarios. For instance, in [64-66] a systematic approach based on a nonlocal generalization of the transformation-optics [67,68] paradigm was proposed. In [49], in order to deal with similar nonlocality issues (angle-dependent surface impedances) in the planar case, a multilayered implementation of the metasurfaces was successfully carried out, based on a general synthesis procedure originally put forward in [69] for the design of computational metamaterials. Here, we explore the generalization of this planar implementation to our cylindrical scenario.

To this aim, as schematized in Figure 8, each of the idealized (i.e., zero-thickness) cylindrical metasurfaces is replaced by a physical structure composed of subwavelength material layers. With a view toward technological feasibility, the material constituents are assumed as homogeneous, isotropic and nonmagnetic, so that the only optimization parameters available are the layer thicknesses and their (suitably constrained) dielectric permittivities. Moreover, we consider a number of four layers as a reasonable tradeoff between response complexity (and, hence, nonlocality-tailoring capabilities) and computational burden (as well as fabrication complexity). Details on the synthesis procedure are given in Appendix A.

As a representative example, we consider a parameter configuration (green-circle marker in Figure 4) with $k_0 R_s = 10, k_0 R_i = 20, k_0 R_1 = 11.40, k_0 R_2 = 14.50$, characterized by a sensibly stronger nonlocality ($F \approx -7\text{dB}$), for which the multilayer-synthesis procedure yields the parameters given in Table 1. We observe that, for both metasurfaces, the multilayered implementation features alternating layers made of lossy negative-permittivity and active positive-permittivity constituents, with total thickness of about $0.25\lambda_0$. Although the emphasis of this prototype study is on a proof-of-concept demonstration and on the illustration of the phenomenology, rather than technological and fabrication-related aspects, the permittivity values are constrained within realistic bounds. For instance, the parameters of the negative-permittivity constituents are consistent with those of plasmonic materials (e.g., transparent conductive oxides [70]) at optical wavelengths, and the level of gain is comparable with those attainable in quantum-dot-based active media [62,71].

Figure 9 shows the corresponding results. As it can be expected, the ideal modal surface-impedances [Figure 9(a) and Figure 9(b)] now exhibit more significant variations (i.e., more pronounced nonlocality). As a source-field, we consider the same diffraction-limited, real-valued profile as in Figure 5. To provide an independent validation, the field imaged by the multilayered structure is now computed via finite-element-based numerical simulations (see Appendix B for details). As we can observe, the agreement with the source-field profile is excellent for the (dominant) real part, with some residual oscillations around the ideally zero imaginary part. It is also interesting to observe that, for this parameter configuration, the local approximation [with $\bar{Z}_1 = (0.017 - i0.056)\eta_0$ and $\bar{Z}_2 = -(0.022 + i0.080)\eta_0$, from (20)], which is used here as a reference case like the vacuum case was used in Figure 5 and Figure 7, yields remarkably poorer results, as a further confirmation of the non-negligible nonlocal effects.

Overall, the four-layer optimized geometry works reasonably well for the moderate degree of required nonlocality that we have considered in this example. This demonstrates that, even for designs involving non-negligible levels of nonlocality, the metasurface implementation remains fairly simple, in terms of few cylindrical layers, without requiring extreme-parameter media. Clearly, more complex designs, featuring additional degrees of freedom, may be necessary in order to capture stronger desired nonlocal responses.

**C. Stability analysis**

The results illustrated above are derived through a time-harmonic wave-scattering formalism, and therefore pertain to the *steady-state* response of the system. However, the presence of gain in our non-Hermitian configuration can give rise to optical instability; in other words, the structure may support self-oscillations. In [49], with reference with the PT-symmetric planar counterpart, it was shown that the system can be made *unconditionally stable* by suitably choosing the dispersion of the two metasurfaces. A similar analysis is carried out here for our cylindrical scenario. To this aim, we revisit the scattering model in (4)-(11), but now assuming the two surface impedances $Z_1$ and $Z_2$ as known terms, and the coefficients $A_1^{(n)}$ and $A_3^{(n)}$ as unknowns. By solving the resulting linear system of equations, we can calculate the scattering parameters (transfer-functions)

$$T_n(\omega) = \frac{A_3^{(n)}\left[Z_1(\omega), Z_2(\omega)\right]}{A_1^{(n)}\left[Z_1(\omega), Z_2(\omega)\right]}, \qquad (21)$$

which relate the angular-momentum modes at the source and image surfaces [cf. (7) and (8)]. The analytical expressions of the transfer functions in (21) are provided in Appendix C. In (21), only the frequency dependence is explicitly highlighted, and related to the dispersion models of the surface impedances $Z_1(\omega)$ and $Z_2(\omega)$. Such dispersive models are constrained by causality (namely, they respect Kramers-Kronig-like equations), but their specific details inherently depend on the physical metasurface implementation. Along the lines of [49], here we consider a simple metasurface implementation in terms of a thin, subwavelength cylindrical layer made of a homogeneous, isotropic, nonmagnetic material characterized by a causal dispersion law. The resulting dispersive model, detailed in Appendix C, is especially suited for the quasi-local implementation illustrated in Sec. IV-A. Accordingly, as a representative example, we consider the parameter configuration in Figure 7.

Figure 10 shows an example of causal dispersive laws (see Appendix C for details) for the two surface impedances, obtained by enforcing at a center angular frequency $\omega_c$ the corresponding average values [from (20)],

$$Z_1(\omega_c) = \bar{Z}_1 = (0.557 + i0.205)\eta_0, \tag{22}$$

$$Z_2(\omega_c) = \bar{Z}_2 = (-0.690 + i0.103)\eta_0. \tag{23}$$

For this configuration, Figure 11 shows the transfer-functions (magnitude) in (21), over the complex $\omega$-plane, for the ($0 \leq n \leq 5$) angular-momentum modal orders relevant to the example in Figure 7. As it can be qualitatively observed, the responses are only weakly dependent on the modal order, and all poles are confined to the lower half of the complex plane $\text{Im}(\omega) < 0$, which, in view of the implied $\exp(-i\omega t)$ time-harmonic convention, guarantees that the system is *unconditionally stable* for any temporal excitation. More in detail, we numerically verified the presence of poles at $\omega \approx (0.83 - i0.6)\omega_c$, with variations on the second significant digits depending on the angular-momentum modal order $n$. However, it is worth stressing that different parameter choices in the dispersion laws, as well as in the lens configuration, may give rise to transfer functions exhibiting poles in the upper half-plane $\text{Im}(\omega) > 0$, thereby driving the system to an unstable (self-radiating) regime. Therefore, care should be exerted in ascertaining the stability on a case-by-case basis. Overall, the main indications emerged for the planar scenario [49] remain valid for the cylindrical geometry considered here.

**D. Remarks**

In connection with the practical feasibility of our proposed configuration, it should be stressed that the infinite extent (along the *z*-direction) assumed in our study is not a relevant constraint. In fact, in view of the assumed *z*-directed electric field, the structure can be longitudinally truncated with two perfectly electric conducting (PEC) parallel planar walls (in the *x-y* plane) placed at an arbitrary distance, without affecting the validity of our in-plane mathematical formulation. In this case, the excitation given by (3) may be mimicked by a sheet distribution of axial electrical currents enforced at the source surface ($r = R_s$) and extending up to the PEC walls. At microwave frequencies, where the PEC condition is well approximated by metals, this parallel-plate waveguide setup represents a typical implementation of 2-D metamaterial configurations. At optical frequencies, where metals behave quite differently, a PEC-like truncation condition may be attained by using photonic crystals operating in the bandgap. In a practical scenario, provided that the cylinder is sufficiently long, the imaging properties described here would hold far enough from the edges of the finite structure.

Another relevant question is whether the inherently nonlocal and non-Hermitian (with gain and loss regions) nature of our proposed configuration poses insurmountable challenges for its realization. Within this framework, it makes sense to compare our proposed configuration with alternative metamaterial strategies for imaging and magnification, based on transformation optics [5,6], negative refraction [7-9] or hyperbolic dispersion [15-17]. From the implementation viewpoint, all these strategies eventually rely on multilayered structures featuring thin material layers and/or resonating elements. In addition, transformation-optics-based strategies typically require complex spatial tailoring of the constitutive parameters. By comparison, in its simplest quasi-local configuration, our proposed strategy can be implemented via only two thin material layers. Even in those cases where nonlocality is non-negligible, each metasurface can be implemented via only few material layers (e.g., four, in the case considered in Figure 9). Therefore, we argue that, in terms of structural complexity, our implementation is certainly comparable with the above alternatives; in fact, it could become even simpler than the alternatives for those parameter configurations featuring weak nonlocality.

In connection with the inherent non-Hermitian character of our proposed design and, in particular, the presence of gain, it is worth stressing that a comparison with *idealized* (lossless) metamaterial alternatives would be quite unfair. In fact, for all the aforementioned alternative strategies, the inevitable presence of losses substantially curtails the resolution and transmittance [18,19]. Transformation-optics-based designs leading to negative-permittivity and negative-permeability media [5,6] are particularly sensitive to the detrimental effects of losses [72,73]. As previously mentioned, a possible strategy to overcome these effects is to introduce gain-material constituents so as to compensate for losses [24,25].

In our design, loss and gain are not considered as second-order effects to compensate for. Instead, they are contemplated from the very beginning, and their tailored interplay is instrumental to attain the desired functionality. Therefore, we argue that our proposed non-Hermitian design is comparable with loss-compensated alternative implementations in terms of realization complexity, and it broadens the conventional framework of gain-induced effects in metamaterials.

We also emphasize that our proposed design can readily be extended to other kinds of waves, such as acoustics, for which large levels of gain/amplification are easier to attain (see, e.g., [50]). A final remark is related to the issue of bandwidth: while passive metamaterials are fundamentally limited by constraints on their frequency dispersion stemming from Kramers-Kronig relations for passive media, and therefore the unusual imaging properties of negative-index or transformation-optics lenses are typically limited to a narrow range of frequencies, active metamaterials may overcome these limitations. While a detailed study on the bandwidth performance of the proposed imaging system is beyond the scope of this paper, and the stability issues mentioned above may fundamentally limit the overall achievable bandwidth of a practical device, it is expected that the bandwidth of operation of the proposed loss-gain cylindrical lens may be superior to the one of metamaterial devices based on only passive elements.

## V. CONCLUSIONS AND OUTLOOK

In conclusion, we have shown that magnified, diffraction-limited imaging with reduced aberrations can be attained by means of a cylindrical lensing system relying on a pair of non-Hermitian, nonlocal metasurfaces. Within certain parameter ranges, nonlocality can be suitably mitigated so that fairly good results can be achieved by employing local metasurfaces. For the more general case, we have demonstrated a multilayered implementation whose nonlocal response can be tailored so as to approximately capture the idealized response. We have also addressed the relevant issue of stability, showing that metasurface dispersion laws can be chosen in such a way to render the system unconditionally stable for any temporal excitation.

The above results complement and expand the previous study in [49] on PT-symmetric planar lenses, and set the stage for the development of a rather general platform for field-manipulation and processing, not necessarily restricted to electromagnetics. Within this framework, current and future investigations are aimed at exploring more in detail the implementation-related issues, in terms of specific material constituents and sensitivity to fabrication tolerances. Also of great interest are possible extensions of the field-manipulation capabilities, along the lines of the computational-metamaterial paradigm introduced in [69].

**ACKNOWLEDGMENTS**

This work was partially supported by the Air Force Office of Scientific Research with grant No. FA9550-13-1-0204, the Office of Naval Research with grant No. N00014-15-1-2685, the Simons Foundation and the Welch Foundation with grant No. F-1802.

**APPENDIX A: DETAILS ON THE MULTILAYERED IMPLEMENTATION**

Assuming the two metasurfaces in Figure 1 implemented by multilayered structures as schematized in Figure 8, the electromagnetic response can still be calculated via the general Fourier-Bessel representation in (4) and (5), which needs to be extended in each of the homogeneous, isotropic material layers with proper adjustments in the wavenumber and characteristic impedances. Instead of the impedance boundary conditions in (11), now the electric field continuity must be enforced at each interface. From the computational viewpoint, it is expedient to utilize a transfer-matrix method [74,75], which allows to systematically relate the field-expansion coefficients at the two ends (source and image) of the structure. By maintaining the same notation as in Sec. III-A, we keep referring to $A_1^{(n)}$, $B_1^{(n)}$ and $A_3^{(n)}$ as the expansion coefficients at the source and image surfaces, respectively, with $B_3^{(n)} = 0$ due to the radiation condition. Unlike the idealized impedance-surface-based synthesis in Sec. III-A, we are now interested in synthesizing the actual multilayers, in terms of the layer thicknesses and permittivities. This renders the problem *nonlinear*, and not solvable analytically in closed form. In other words, the impedance-matching condition at the source-surface [cf. (6)] as well as the magnified-imaging condition [cf. (9)] can no longer be enforced analytically, but only in a *weak* fashion. Accordingly, we define a cost function

$$J(\underline{\varepsilon}_r,\underline{d})=\frac{1}{N+1}\sum_{n=0}^{N}\left|\frac{A_3^{(n)}H_n^{(1)}(k_0R_i)-A_1^{(n)}H_n^{(1)}(k_0R_s)}{A_1^{(n)}H_n^{(1)}(k_0R_s)}\right|^2+\left|B_1^{(n)}\right|^2, \quad (A1)$$

whose global minimum (zero) corresponds to the exact enforcement of the above conditions [cf. (6) and (9)]. In (A1), $\underline{\varepsilon}_r$ and $\underline{d}$ compactly denote two arrays embedding the relative permittivities and thicknesses of the two multilayers implementing the metasurfaces, which constitute the optimization parameters in our procedure. Due to the aforementioned nonlinear character of the problem, the cost function in (A1) is likely to exhibit many local minima (corresponding to false solutions), and there is no guarantee that the optimization procedure will converge to the sought global minimum.

Our optimization strategy is similar to that successfully utilized in [49,69], and relies on a standard Nelder-Mead (downhill-simplex) unconstrained minimization algorithm implemented in the Matlab optimization toolbox [76]. To ensure that the search-space is adequately explored, we randomly move the initial guess across a reasonably broad parameter range. Moreover, we enforce some feasibility-related constraints on the optimization parameters. In particular, besides the aforementioned non-magnetic character, we constrain the positive real-part of the relative permittivities within the range $1\leq \text{Re}(\varepsilon_r)\leq 10$, with the negative imaginary part (representative of gain) restricted as $-\text{Im}(\varepsilon_r)\leq 0.1\text{Re}(\varepsilon_r)$. No explicit constraint is assumed on the negative real-part as well as on the positive imaginary-part (representative of losses) of the permittivities, but negative-permittivity constituents are constrained to be lossy. The thicknesses are constrained within the range $[0.015\lambda_0, 0.1\lambda_0]$. The above constraints are enforced in a *soft* fashion, by suitably choosing the initial-guess parameter ranges, and discarding *a posteriori* those candidate solutions falling outside the allowed ranges.

While the convergence to the global minimum cannot be guaranteed, we found that, for moderate degrees of nonlocality, the above strategy typically led to reasonably low levels of the cost function ($\sim 0.05$), corresponding to satisfactorily good imaging accuracy. However, we also found some particularly high degrees of nonlocality that are not accurately captured by the four-layer structures, and might require more complex implementations.

**APPENDIX B: DETAILS ON THE NUMERICAL SIMULATIONS**

The field distributions pertaining to the idealized (zero-thickness metasurface) structures are computed via the Fourier-Bessel series in (4), while the one pertaining to the multilayered implementation is computed via the finite-element-based commercial software COMSOL Multiphysics [77]. In this case, the structure is excited by enforcing at the source surface $r = R_s$ the field distribution computed via (7), and a perfectly matched layer is used as a termination, in order to avoid fictitious scattering. The structure is discretized by means of an adaptive discrete mesh, with size ranging from a minimum of $\lambda_0 / 300$ (in the thin material layers) to a maximum of $\lambda_0 / 60$ (in the vacuum regions), resulting into about 3 million degrees of freedom. The MUMPS direct solver (with default parameters) is utilized.

**APPENDIX C: DETAILS ON THE STABILITY ANALYSIS**

The transfer-functions in (21) are computed by solving the linear system of equations arising from (10) and (11) [with (6)], by assuming the surface impedances $Z_1$ and $Z_2$ as known terms, and the coefficients $A_1^{(n)}$ and $A_3^{(n)}$ as unknowns. We obtain

$$T_n = \frac{A_3^{(n)}}{A_1^{(n)}} = \frac{a_0}{a_1^{(n)} + a_2^{(n)} + a_3^{(n)}}, \tag{C1}$$

with

$$a_0 = \frac{16 Z_1 Z_2}{\pi^2 k_0^2 R_1 R_2}, \tag{C2}$$

$$\begin{aligned} a_1^{(n)} &= Z_2 \dot{H}_n^{(1)}(k_0 R_2) H_n^{(2)}(k_0 R_2) \\ &\times \left\{ Z_1 H_n^{(1)}(k_0 R_1) \dot{H}_n^{(2)}(k_0 R_1) - H_n^{(2)}(k_0 R_1) \left[ Z_1 \dot{H}_n^{(1)}(k_0 R_1) + i\eta_0 H_n^{(1)}(k_0 R_1) \right] \right\}, \end{aligned} \tag{C3}$$

$$\begin{aligned} a_2^{(n)} &= H_n^{(1)}(k_0 R_2) \left[ \eta_0 H_n^{(2)}(k_0 R_2) + i Z_2 \dot{H}_n^{(2)}(k_0 R_2) \right] \\ &\times \left\{ H_n^{(1)}(k_0 R_1) \left[ \eta_0 H_n^{(2)}(k_0 R_1) + i Z_1 \dot{H}_n^{(2)}(k_0 R_1) \right] - i Z_1 \dot{H}_n^{(1)}(k_0 R_1) H_n^{(2)}(k_0 R_1) \right\}, \end{aligned} \tag{C4}$$

$$a_3^{(n)} = -\left[ \eta_0 H_n^{(1)}(k_0 R_2) H_n^{(2)}(k_0 R_1) \right]^2, \tag{C5}$$

where the frequency-dependence in the surface impedances is omitted for notational compactness, the overdot denotes differentiation with respect to the argument, and all other symbols have already been defined.

The dispersive model of the surface impedances is derived assuming a physical implementation in terms of a thin, subwavelength cylindrical layer made of a homogeneous, isotropic, nonmagnetic material with relative permittivity $\varepsilon_r$. For illustration, we can refer to the schematic in Figure 8, assuming only one layer of thickness $d$, extending over the annular region $R_- < r < R_+$, with $R_- = R - d/2$, $R_+ = R + d/2$, and $R$ denoting the nominal radial position of the ideal metasurface. By solving the arising scattering problems, and matching (in the limit $d \ll \lambda_0$) the transmission coefficient relating the angular-momentum modal orders at the surfaces $r = R_-$ and $r = R_+$ with that obtained in the presence of an ideal metasurface at $r = R$, we can calculate the equivalent surface impedance

$$Z = \frac{b_0^{(n)} \eta_0}{k_0 \dot{H}_n^{(2)}(k_0 R_-) b_1^{(n)} + k H_n^{(2)}(k_0 R_-) b_2^{(n)}}, \tag{C6}$$

with

$$b_0^{(n)} = \frac{16 k_0 R H_n^{(1)}(k_0 R) H_n^{(2)}(k_0 R)}{\pi (4R^2 - d^2)}, \tag{C7}$$

$$\begin{aligned}
b_1^{(n)} &= H_n^{(1)}(kR_+) \left[ k H_n^{(1)}(k_0 R_-) \dot{H}_n^{(2)}(kR_+) - k_0 \dot{H}_n^{(1)}(k_0 R_+) H_n^{(2)}(kR_-) \right] \\
&+ H_n^{(1)}(kR_-) \left[ k_0 \dot{H}_n^{(1)}(k_0 R_+) H_n^{(2)}(kR_+) - k H_n^{(1)}(k_0 R_+) \dot{H}_n^{(2)}(kR_+) \right] \\
&+ k \dot{H}_n^{(1)}(kR_+) \left[ H_n^{(1)}(k_0 R_-) H_n^{(2)}(kR_-) - H_n^{(1)}(k_0 R_-) H_n^{(2)}(kR_+) \right],
\end{aligned} \tag{C8}$$

$$\begin{aligned}
b_2^{(n)} &= \dot{H}_n^{(2)}(kR_+) \left[ k H_n^{(1)}(k_0 R_+) \dot{H}_n^{(1)}(kR_-) - k_0 \dot{H}_n^{(1)}(k_0 R_-) H_n^{(1)}(kR_+) \right] \\
&+ \dot{H}_n^{(2)}(kR_-) \left[ k_0 \dot{H}_n^{(1)}(k_0 R_+) H_n^{(1)}(kR_+) - k H_n^{(1)}(k_0 R_+) \dot{H}_n^{(1)}(kR_+) \right] \\
&+ k_0 H_n^{(2)}(kR_+) \left[ \dot{H}_n^{(1)}(k_0 R_-) \dot{H}_n^{(1)}(kR_+) - \dot{H}_n^{(1)}(k_0 R_+) \dot{H}_n^{(1)}(kR_-) \right],
\end{aligned} \tag{C9}$$

where $k = k_0 \sqrt{\varepsilon_r}$.

We observe that the expression in (C6) depends on the angular-momentum modal order $n$. However, for the assumed parameter configuration (as in Figure 7), such dependence is quite mild. We verified that by approximating the equivalent surface impedance via local (tangent-plane) application of the expression utilized in [49] planar case,

$$Z \approx \frac{1}{i \omega d (1 - \varepsilon_r) \varepsilon_0}, \tag{C10}$$

we obtain a reasonably ($\leq 8\%$) small error over the parameter range of interest.

To introduce a causal dispersion law, we assume for the passive metasurface ($Z_1$) a Lorentz-type dispersion model for the material layer,

$$\varepsilon_{r1}(\omega) = \varepsilon_{r1}^{(\infty)} - \frac{\omega_{p1}^2}{\omega^2 - \omega_{01}^2 + i\Gamma_1\omega}, \tag{C11}$$

whereas for the active metasurface ($Z_2$) we consider anti-Lorentz dispersion model,

$$\varepsilon_{r2}(\omega) = \varepsilon_{r2}^{(\infty)} + \frac{\omega_{p2}^2}{\omega^2 - \omega_{02}^2 + i\Gamma_2\omega}. \tag{C12}$$

The dispersion laws in Figure 10 are obtained assuming $d = \lambda_c/20$ (with $\lambda_c = 2\pi c/\omega_c$ denoting the vacuum wavelength at the center frequency) and a parameter configuration in (C11) and (C12) (given in the figure caption) that satisfies the nominal-design conditions in (22) and (23) at the center angular frequency $\omega_c$. Clearly, given the number of adjustable parameters in (C11) and (C12), there are infinite parameter configurations that would yield the same desired impedance values at $\omega_c$. However, the stability of the system is not always guaranteed, and should be independently assessed for each case.

[75] C. A. Valagiannopoulos and P. Alitalo, Phys. Rev. B **85**, 115402 (2012).
[76] Matlab Optimization Toolbox reference guide, available at: www.mathworks.com/products/optimization/index.html
[77] COMSOL Group, *COMSOL Multiphysics: Version 5.1* (COMSOL, Stockholm, 2015)30

Table 1. Synthesis parameters for the multilayered implementation of the nonlocal metasurfaces (see the schematic in Figure 8), for $k_0 R_s = 10$, $k_0 R_i = 20$, $k_0 R_1 = 11.4$, $k_0 R_2 = 14.5$, and $N = 5$.

|       | Metasurface 1 |           | Metasurface 2 |           |
|-------|---------------|-----------|---------------|-----------|
| Layer | $\varepsilon_r$ | $d/\lambda_0$ | $\varepsilon_r$ | $d/\lambda_0$ |
| 1     | $4.665 - i0.4658$  | 0.096 | $5.139 - i0.504$  | 0.089 |
| 2     | $-0.1402 + i0.379$ | 0.032 | $-0.727 + i0.325$ | 0.042 |
| 3     | $5.179 - i0.518$   | 0.100 | $3.917 - i0.391$  | 0.097 |
| 4     | $-0.523 + i0.089$  | 0.022 | $-0.377 + i0.228$ | 0.030 |

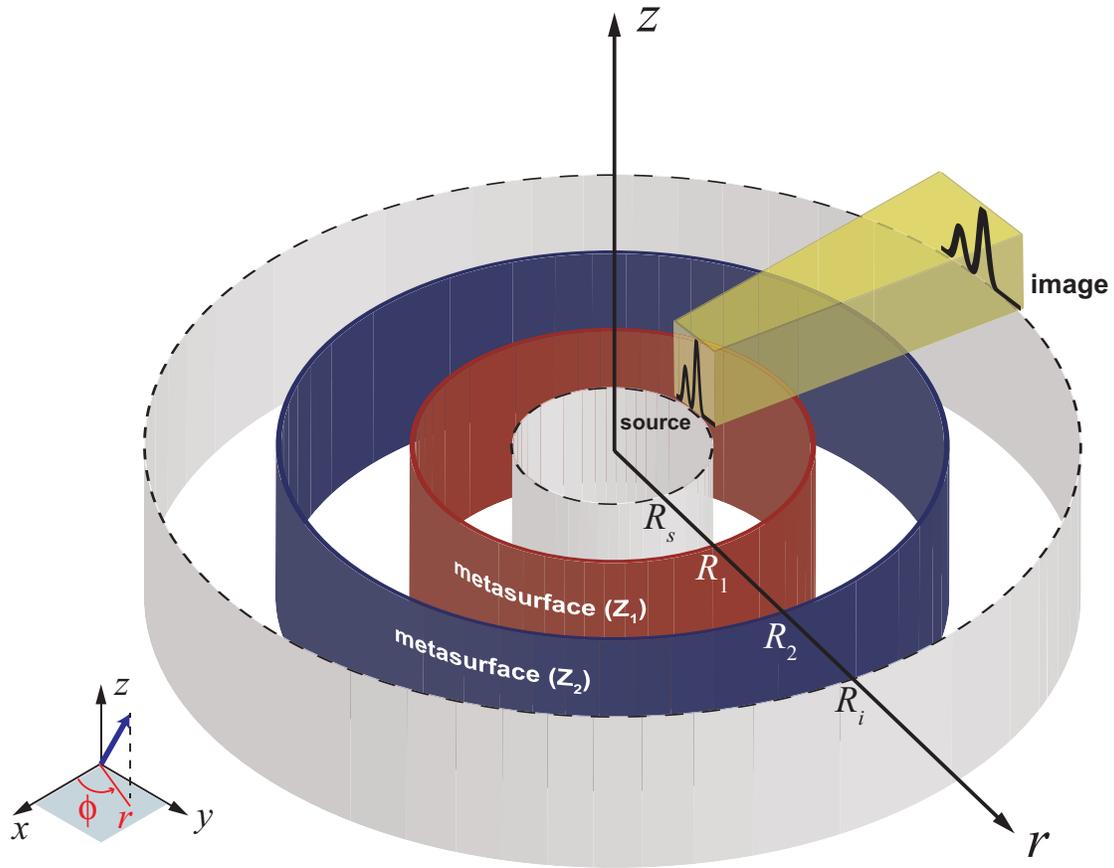

Figure 1. Problem schematic: A cylindrical lensing system composed of two concentric metasurfaces of radii $R_1$ and $R_2$ and impedances $Z_1$ and $Z_2$, respectively, embedded in vacuum. An assigned field distribution at a surface of radius $R_s < R_1$ is imaged at a surface of radius $R_i > R_2$, thereby attaining a geometrical magnification by a factor $R_i/R_s > 1$. Also shown are the associated Cartesian and cylindrical references systems. Geometry and field quantities are assumed as invariant along the $z$-direction.

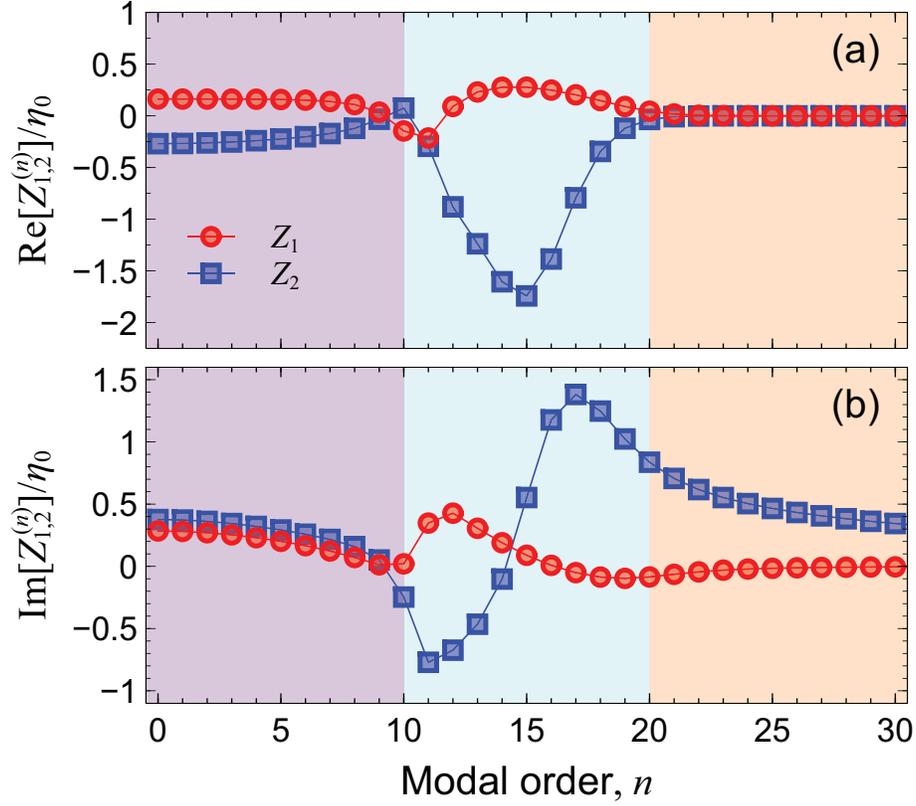

Figure 2. Geometry as in Figure 1, with $k_0 R_s = 10$, $k_0 R_i = 20$, $k_0 R_1 = 13$, and $k_0 R_2 = 17$. (a), (b) Real and imaginary part, respectively, of the surface impedances $Z_1^{(n)}$ (red circles) and $Z_2^{(n)}$ (blue squares), numerically computed from (12) and (13), respectively, as a function of the angular-momentum order $n$ [in view of the symmetry condition (15), only $n \geq 0$-orders are displayed]. Values are normalized with respect to the vacuum characteristic impedance $\eta_0$. Continuous curves are guides to the eye only.

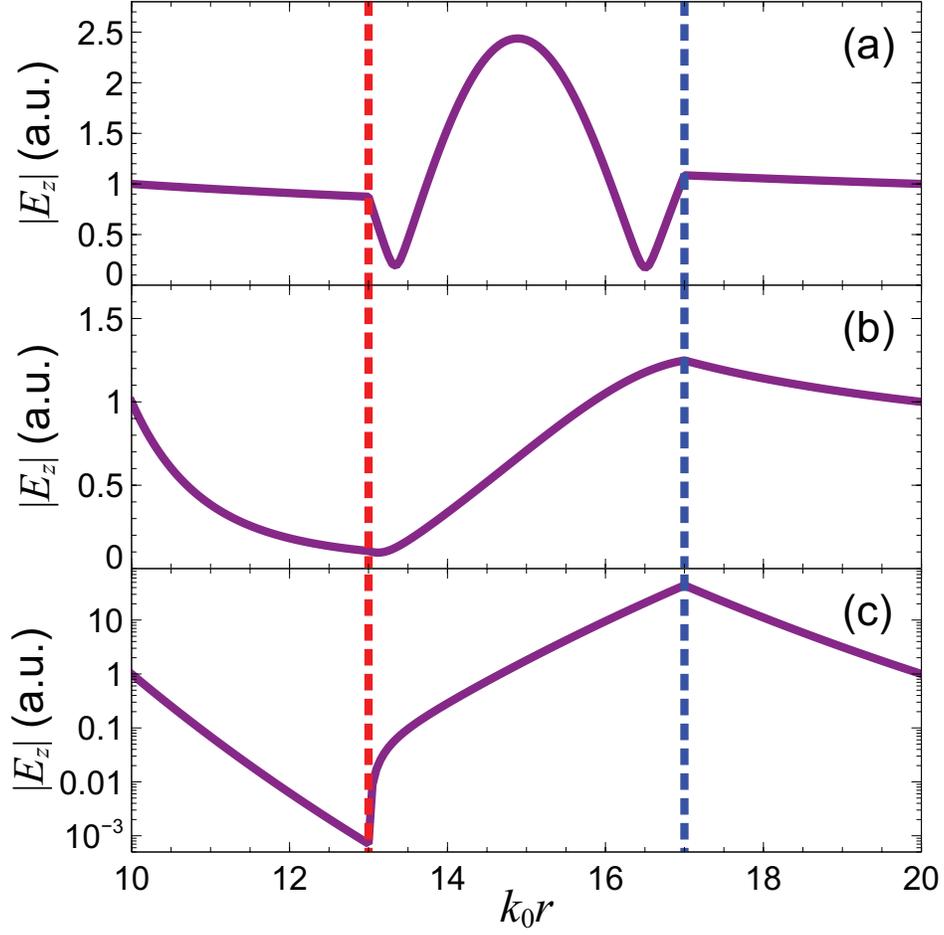

Figure 3. (a), (b), (c) Electric-field (magnitude) radial distributions pertaining to the angular-momentum modal orders $n=2$, $n=15$, and $n=30$, respectively, for the parameter configuration in Figure 2. Field values are normalized with respect to amplitudes at the source surface ($r=R_s$). The red and blue dashed vertical lines indicate the locations of the metasurfaces. The corresponding surface-impedance values [from (12) and (13)] are $Z_1^{(2)}=(0.162+i0.270)\eta_0$, $Z_2^{(2)}=(-0.262+i0.362)\eta_0$, $Z_1^{(15)}=(0.277+i0.089)\eta_0$, $Z_2^{(15)}=(-1.742+i0.553)\eta_0$, $Z_1^{(30)}=-i0.004\eta_0$, $Z_2^{(30)}=i0.344\eta_0$.

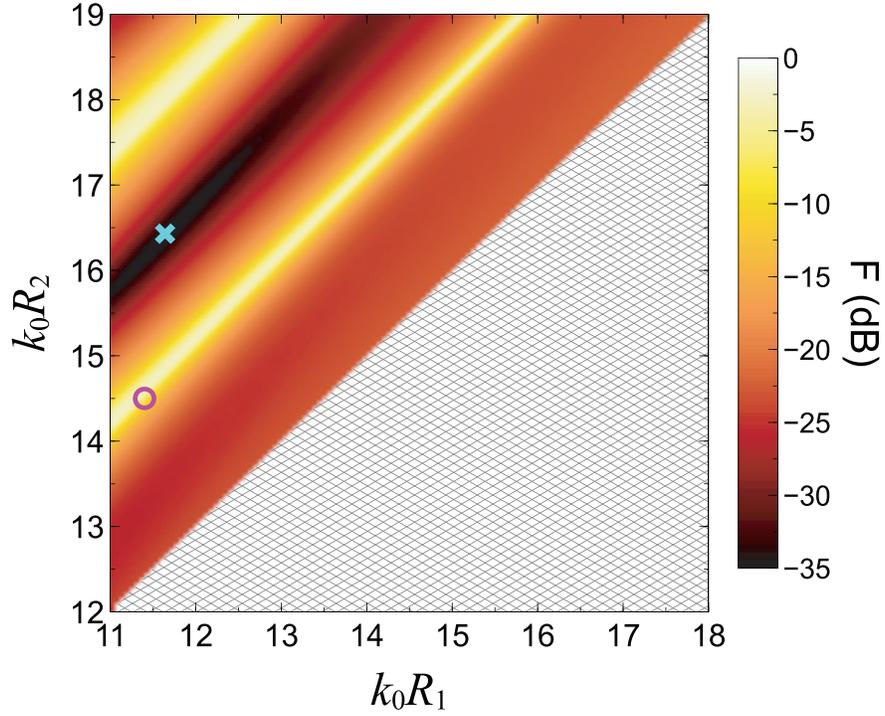

Figure 4. Nonlocality indicator in (19) (in dB scale) for $k_0 R_s = 10$, $k_0 R_i = 20$, $N = 5$, as a function of $k_0 R_1$ and $k_0 R_2$. The cyan-cross marker indicates the parameter configuration $k_0 \hat{R}_1 = 11.64, k_0 \hat{R}_2 = 16.43$, which minimizes the nonlocality ($F \approx -35 \text{dB}$). The green-circle marker indicates the parameter configuration $k_0 R_1 = 11.40, k_0 R_2 = 14.50$, which yields a sensibly stronger nonlocality ($F \approx -7 \text{dB}$). The study is restricted to metasurface distances $k_0 (R_2 - R_1) > 1$, thereby excluding the grey-motif region.

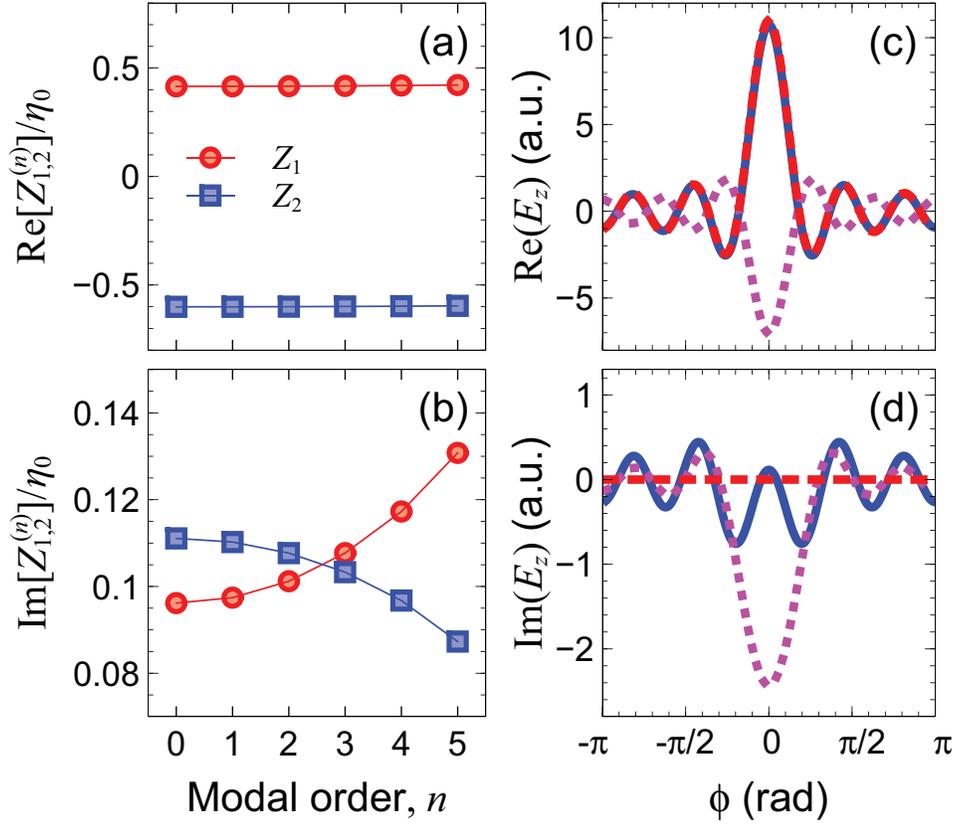

Figure 5. Geometry as in Figure 1, with $k_0 R_s = 10$, $k_0 R_i = 20$, $k_0 R_1 = 11.64$, $k_0 R_2 = 16.43$. (a), (b) Real and imaginary part, respectively, of the surface impedances $Z_1^{(n)}$ (red circles) and $Z_2^{(n)}$ (blue squares), numerically computed from (12) and (13), respectively, as a function of the angular-momentum order $n$, up to $N = 5$ [in view of the symmetry condition (15), only $n \geq 0$-orders are displayed]. Values are normalized with respect to the vacuum characteristic impedance $\eta_0$. Continuous curves are guides to the eye only. (c), (d) Real and imaginary parts, respectively, of the enforced source-field profile in (7) (with $A_1^{(n)} = 1$ for $n = -5,..,5$, and $A_1^{(n)} = 0$ otherwise; red-dashed curves), compared with the imaged field profile [computed via (8); blue-solid curves] obtained via the local approximation in (20), yielding constant values of the surface impedances $\bar{Z}_1 = (0.418 + i0.102)\eta_0$ and $\bar{Z}_2 = (-0.599 + i0.109)\eta_0$. Also shown, as a reference (magenta-dotted curves), is the field profile at the image surface in the absence of the cylindrical lens ($\bar{Z}_{1,2} \to \infty$).

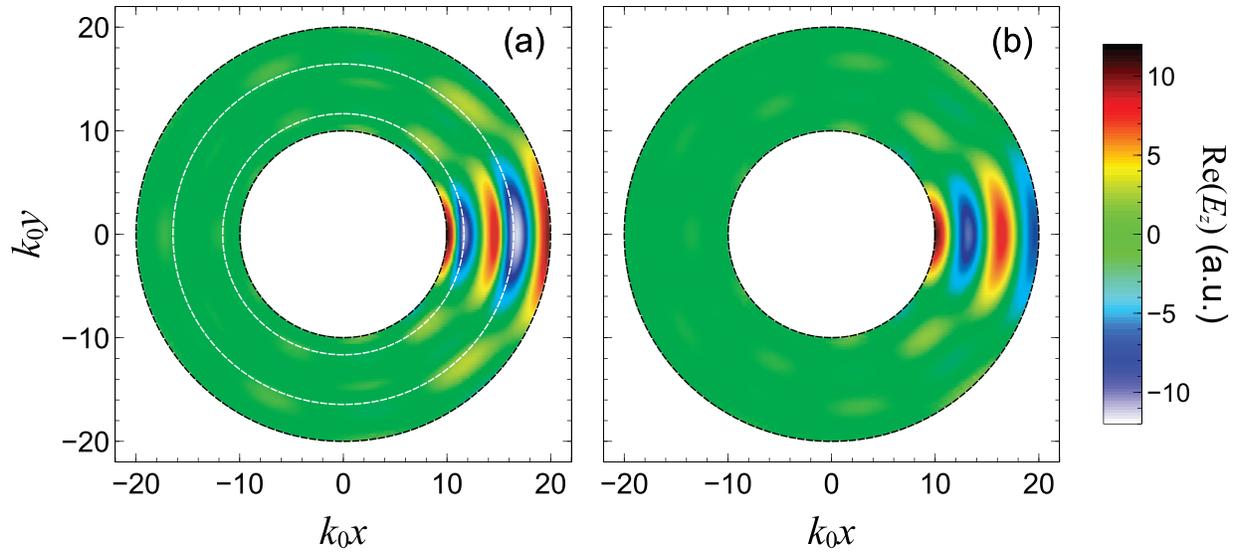

Figure 6. Geometry and parameters as in Figure 5. (a), (b) Real-part field distributions over the entire cylindrical lens domain, in the presence and absence of the metasurfaces, respectively. The black-dashed indicate the locations of the source and image surfaces, while the white-dashed circles indicate the locations of the two metasurfaces.

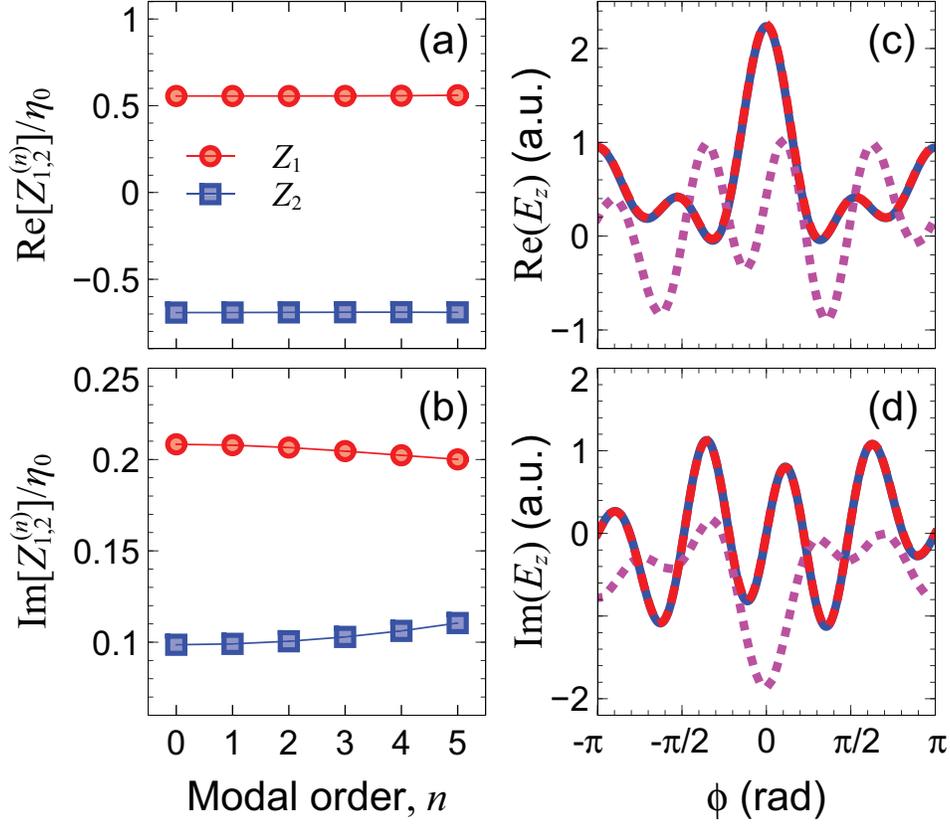

Figure 7. Geometry as in Figure 1, with $k_0 R_s = 13$, $k_0 R_i = 18$, $k_0 R_1 = 13.40$, $k_0 R_2 = 15.65$. (a), (b) Real and imaginary part, respectively, of the surface impedances $Z_1^{(n)}$ (red circles) and $Z_2^{(n)}$ (blue squares), numerically computed from (12) and (13), respectively, as a function of the angular-momentum order $n$, up to $N = 5$ [in view of the symmetry condition (15), only $n \geq 0$ - orders are displayed]. Values are normalized with respect to the vacuum characteristic impedance $\eta_0$. Continuous curves are guides to the eye only. (c), (d) Real and imaginary parts, respectively, of the enforced source-field profile in (7) (with $A_1^{(-5)} = A_1^{(-2)} = A_1^{(1)} = A_1^{(3)} = 0.1$, $A_1^{(-4)} = A_1^{(0)} = 0.6$, $A_1^{(-3)} = A_1^{(-1)} = -A_1^{(4)} = 0.2$, $A_1^{(2)} = -A_1^{(5)} = 0.5$, and $A_1^{(n)} = 0$ otherwise; red-dashed curves), compared with the imaged field profile [computed via (8); blue-solid curves] obtained via the local approximation in (20), yielding constant values of the surface impedances $\bar{Z}_1 = (0.557 + i0.205)\eta_0$ and $\bar{Z}_2 = (-0.690 + i0.103)\eta_0$. Also shown, as a reference (magenta-dotted curves), is the field profile at the image surface in the absence of the cylindrical lens ($\bar{Z}_{1,2} \to \infty$).

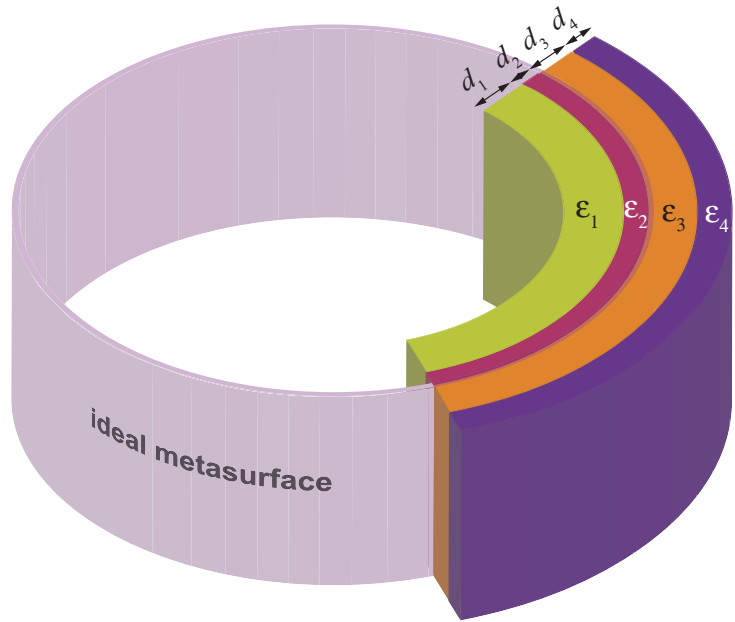

Figure 8. Schematic of the multilayered implementation. Each of the idealized (i.e., zero-thickness) metasurfaces is implemented as a physical structure composed of four layers of homogeneous, isotropic, nonmagnetic material constituents, with subwavelength thicknesses $d_j$ and relative permittivites $\varepsilon_j$, $j=1,2,3,4$.

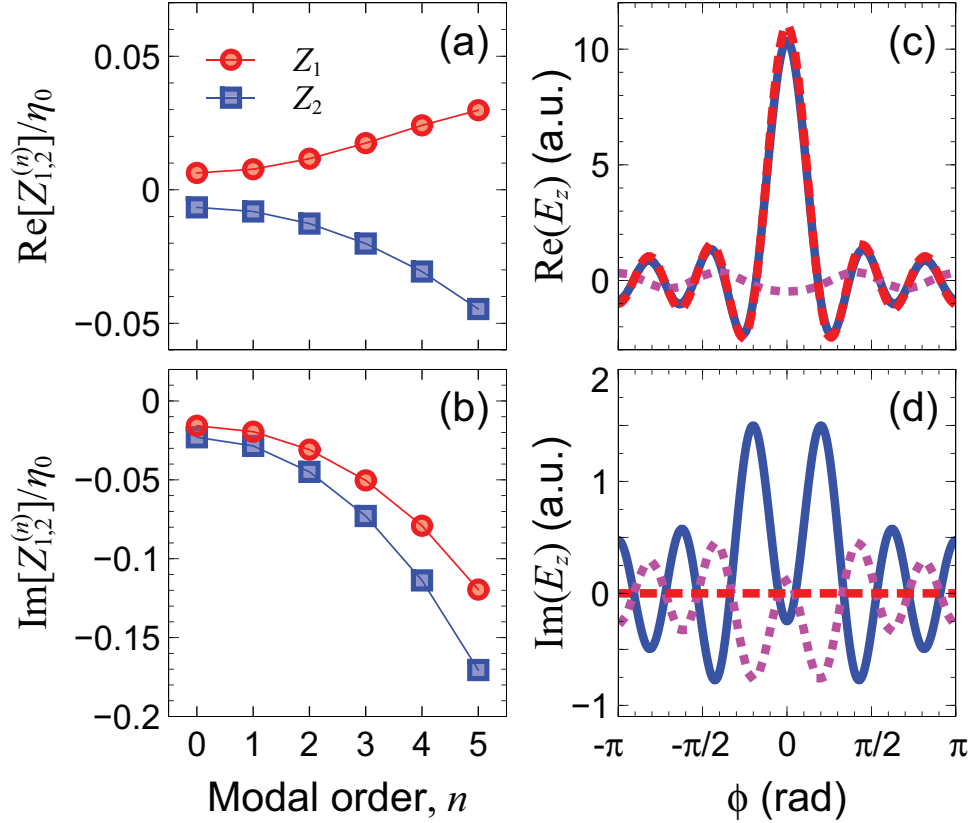

Figure 9. (a) Geometry as in Figure 1, with $k_0 R_s = 10$, $k_0 R_i = 20$, $k_0 R_1 = 11.4$, $k_0 R_2 = 14.5$. (a), (b) Real and imaginary part, respectively, of the surface impedances $Z_1^{(n)}$ (red circles) and $Z_2^{(n)}$ (blue squares), numerically computed from (12) and (13), respectively, as a function of the angular-momentum order $n$, up to $N = 5$ [in view of the symmetry condition (15), only $n \geq 0$-orders are displayed]. Values are normalized with respect to the vacuum characteristic impedance $\eta_0$. Continuous curves are guides to the eye only. (c), (d) Real and imaginary parts, respectively, of the enforced source-field profile in (7) [with $A_1^{(n)} = 1$ for $n = -5,..,5$, and $A_1^{(n)} = 0$ otherwise; red-dashed curves], compared with the imaged field profile (blue-solid curves) obtained via multilayered implementation of the metasurfaces (see Figure 8 and Table 1) and computed via finite-element simulations. Also shown, as a reference (magenta-dotted curves), is the imaged field profile obtained via the local approximation in (20), yielding constant values of the surface impedances $\bar{Z}_1 = (0.017 - i0.056)\eta_0$ and $\bar{Z}_2 = -(0.022 + i0.080)\eta_0$.

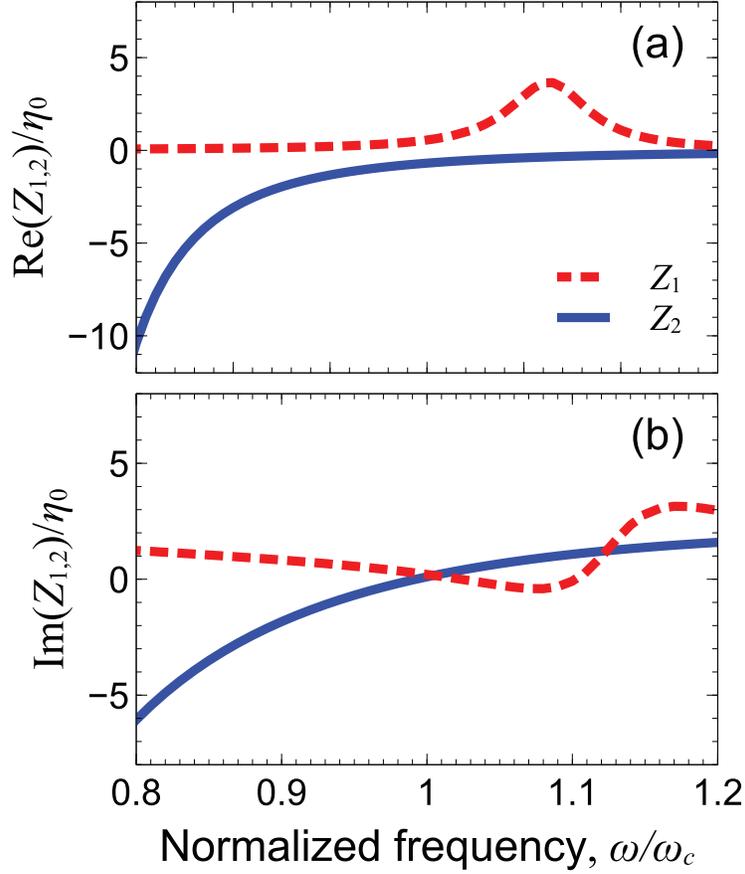

Figure 10. Geometry and parameters as in Figure 7. (a), (b) Real and imaginary part, respectively, of the dispersive laws pertaining to the surface impedances $Z_1$ (red-dashed curves) and $Z_2$ (blue-solid curves). The dispersive models are detailed in Appendix C, and are computed via (C10)-(C12), with $d = \lambda_c/20$, $\varepsilon_{r1}^{(\infty)} = 2.852$, $\varepsilon_{r2}^{(\infty)} = 1.674$, $\omega_{01} = \omega_{02} = \omega_c$, $\omega_{p1} = 0.709\omega_c$, $\omega_{p2} = 0.672\omega_c$, $\Gamma_1 = \Gamma_2 = 0.1\omega_c$; this parameter configuration satisfies the nominal-design conditions in (22) and (23) at the center angular frequency $\omega_c$.

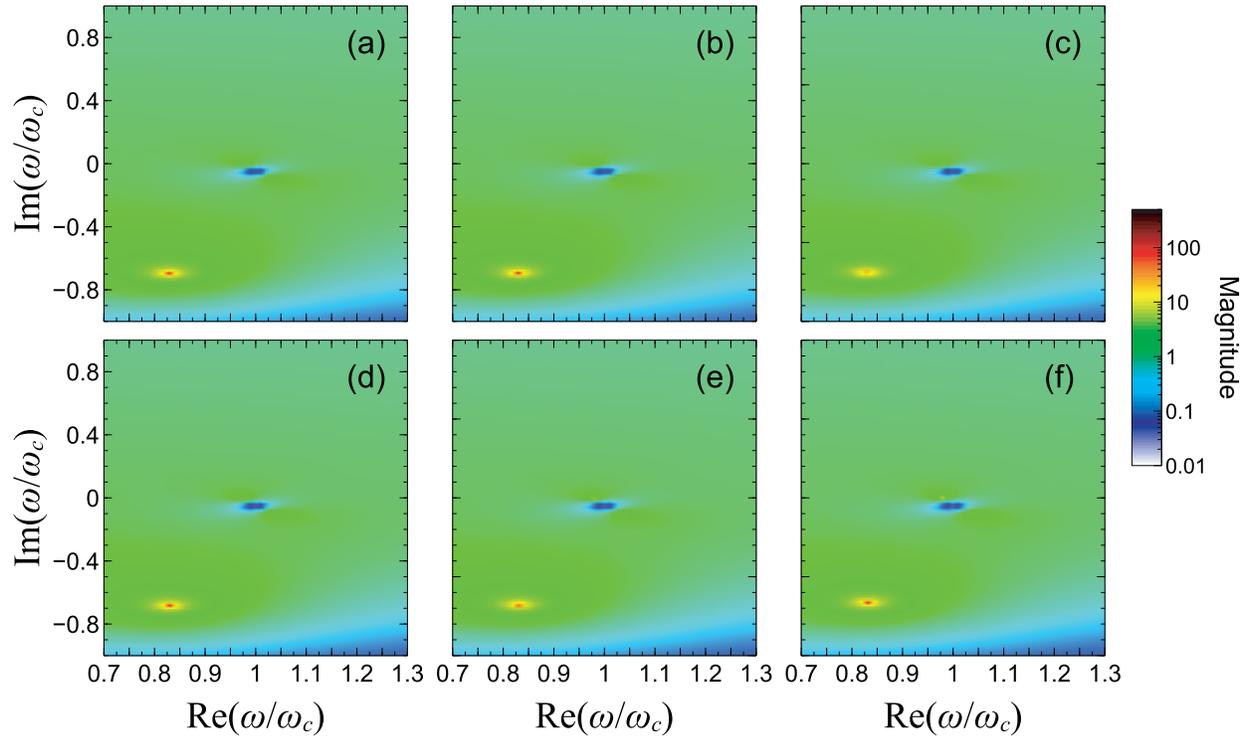

Figure 11. Geometry and parameters as in Figure 7, with surface-impedance dispersion laws as in Figure 10. (a), (b), (c), (d), (e), (f) Magnitude of transfer-functions in (21), $\left|T_n(\omega)\right|$, over the complex $\omega$-plane, for angular-momentum modal orders $n = 0, 1, 2, 3, 4, 5$, respectively. The complex angular frequency is normalized by its center value $\omega_c$, at which the nominal design is attained [see (22) and (23)].